%% file: main.tex
\documentclass[%
 reprint,
 amsmath,amssymb,
 aps,
pr,
]{revtex4-2}

\usepackage[utf8]{inputenc}
\DeclareUnicodeCharacter{05BF}{}
\usepackage{graphicx}
\usepackage{dcolumn}
\usepackage{bm}
\usepackage{hyperref}
\usepackage{gensymb}
\usepackage[utf8]{inputenc}
\usepackage{braket}
\usepackage{amsmath}
\usepackage{color}
\usepackage{soul}
\usepackage{xfrac}
\usepackage{amssymb}
\usepackage{comment}
\usepackage[normalem]{ulem}
\newcommand\redsout{\bgroup\markoverwith{\textcolor{red}{\rule[0.5ex]{2pt}{0.4pt}}}\ULon}

\usepackage{physics}
\usepackage{dsfont}
\usepackage{bbold}

\soulregister\cite{7} 
\soulregister\ref{7} 
\soulregister\eqref{7} 

\begin{document}

\preprint{}

\title{Open transmission channels in multimode fiber cavities with random mode mixing}

\author{Guy Pelc$^{1}$}
\author{Shay Guterman$^{1}$}
\author{Rodrigo Gutiérrez-Cuevas$^{2}$}
\author{Arthur Goetschy$^{2}$}
\author{Sébastien M. Popoff$^{2}$}
\author{Yaron Bromberg$^{1}$}
 \email{yaron.bromberg@mail.huji.ac.il}

\address{$^1$Racah Institute of Physics, The Hebrew University of Jerusalem, Jerusalem 91904, Israel}
\address{$^2$Institut Langevin, ESPCI Paris, PSL University, CNRS, France}

\begin{abstract}
The transport of light in disordered media is governed by open transmission channels, which enable nearly complete transmission of the incident power, despite low average transmission. Extensively studied in diffusive media and chaotic cavities, open channels exhibit unique properties such as universal spatial structure and extended dwell times. However, their experimental study is challenging due to the large number of modes required for control and measurement. We propose a multimode fiber cavity (MMFC) as a platform to explore open channels. Leveraging mode confinement and finite angular spread, MMFCs enabled full channel control, yielding an 18-fold power enhancement in experiment by selectively exciting an open channel with a transmission rate of $0.90 \pm 0.04$. By analyzing 100 transmission matrices of MMFC realizations, we observed a bimodal transmission eigenvalue distribution, indicating high channel control and low losses. The scalability of MMFCs, combined with long dwell times and potential for nonlinear phenomena, offers new opportunities for studying complex wave transport.
\end{abstract}

\maketitle

When coherent light illuminates thick scattering samples, most of the incident power is backscattered. However, theory predicts that with a sufficiently large illumination area, one can identify an incident wavefront that enables all light to be transmitted through the sample via the so-called open transmission channels \cite{dorokhov_coexistence_1984,pendry1990maximal}. Advances in wavefront shaping have allowed precise control over optical wavefronts \cite{vellekoop2007focusing,vellekoop2008universal,mounaix2020time}, enabling the selective excitation of open channels \cite{sarma2016control,bender2020fluctuations}. Over the past decade, several unique properties of open channels have been uncovered, including correspondence with quasi-normal modes \cite{wang_transport_2011,Pea2014,davy2015,davy2018selectively}, universal spatial structure \cite{Choi2011,Davy2015_universal,Ojambati2016,Cao2019_localization,bender2020fluctuations}, and association with extended dwell times inside the sample \cite{shi_statistics_2015,davy2015,durand2019optimizing}.

Increased dwell times in disordered samples enhance light-matter interactions, boost nonlinear effects, and improve environmental sensitivity, key features for sensing applications. Thus, studying the existence of open channels in systems relevant to sensing and nonlinear optics, such as optical fibers, is desirable. 

Transmission through multimode optical fibers shares similarities with scattering samples, as both exhibit random mode coupling \cite{cao2023controlling,bromberg2016control,li2021memory,gutierrez2024characterization}. A key advantage of multimode fibers is that all their modes can be controlled using an SLM \cite{mounaix2019control,kupianskyi2024all}. In standard optical fibers, open channels have little significance since backscattering is negligible. However, by introducing reflective coatings to create a multimode fiber cavity (MMFC), most incident light is backreflected and the concept of open channels becomes applicable. Without mode mixing, an open channel of an MMFC simply corresponds to a guided mode whose propagation constant matches the Fabry–Pérot resonance condition. However, in the presence of strong mode mixing, the existence and properties of open channels ֿare more intricate.

In this Letter, we demonstrate that MMFCs with strong mode mixing can indeed support open channels. We show that despite strong coupling between guided modes, it is possible to selectively excite an open channel and achieve an 18-fold enhancement in transmitted power through the cavity. We find that the transmission eigenvalue distribution exhibits a bimodal shape, reflecting the high control of MMFC channels and its minimal losses. These findings establish MMFCs as a versatile platform for studying and controlling complex optical modes extending beyond open channels, such as recently discovered reflection-less scattering modes \cite{sol2023reflectionless,jiang2024coherent}.

Open channels are found by measuring the transmission matrix of the sample $T$, which relates the input and output fields by $\vec{\psi}_{out}=T \vec{\psi}_{in}$ \cite{popoff_measuring_2010}. The singular value decomposition of $T$, or equivalently, the eigenvectors of the Hermitian matrix $T^\dagger T$, define the transmission eigenchannels of the sample $T^\dagger T\vec{v}_n=\tau_n\vec{v}_n$, where the eigenvalues $\tau_n$ correspond to the transmission rates. To selectively excite the most open channel, the wavefront of the input field is tailored to match the wavefront corresponding to the eigenchannel $\vec{v}_1$ with the highest transmission rate $\tau_1$. The two key ingredients for realizing open channels are, therefore, coherent detection of the light transmitted through the fiber and precise control over the amplitude and phase of the incident wavefront.

\begin{figure}[t!]
 \centering
  \includegraphics[width=1\columnwidth]{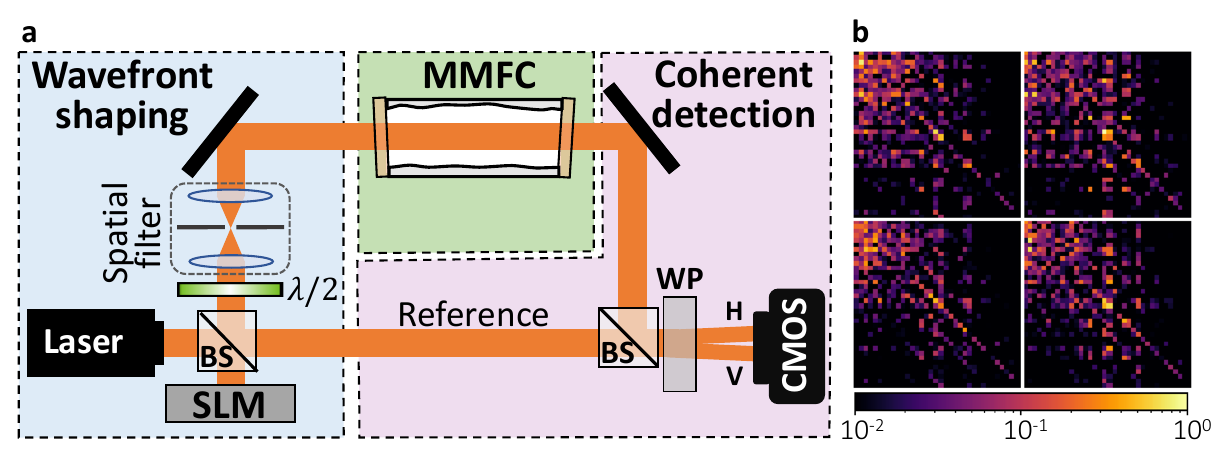}
 \caption{Schematic of the experimental setup and measured transmission matrix. (a) The transmission matrix of the multimode fiber cavity (MMFC) is measured by placing it in one arm of a Mach–Zehnder interferometer. The polarization-dependent complex field at the output of the fiber is measured for a set of input modes excited using the spatial light modulator (SLM). The wavefront required to excite an open channel is computed from the transmission matrix, and then realized using the wavefront shaping apparatus. A spatial filter is used to convert the phase-only SLM to an amplitude and phase modulator. A half-wave plate ($\lambda/2$) is used to control the input polarization, while a Wollaston Prism (WP) before the camera (CMOS) separates horizontal (H) and vertical (V) polarizations. BS: beamsplitter. (b) A measured transmission matrix, presented in the fiber mode basis, exhibiting strong mode mixing. 
The left (right) quadrants correspond to measurements in the horizontal (vertical) input polarizations, while the top (bottom) quadrants correspond to measurements in the horizontal (vertical) output polarizations. In each quadrate, $0.8$ of the energy is concentrated in the off-diagonal elements. 
}
 \label{fig:2}
\end{figure}

To experimentally study the transmission eigenchannels of an MMFC,  we used a one-meter long step-index fiber with a core diameter of $25\,{\mathrm{\mu m}}$ and a numerical aperture of NA = 0.1, coated with reflective coatings with reflectivity $\rho=0.88$. We placed the MMFC in one arm of a Mach-Zehnder interferometer (Fig.~\ref{fig:2}a and Fig.~\ref{fig:exp_setup_extended}). A tunable laser ($\lambda=632\, \mathrm{nm}$) with a long coherence length ($>$ 100 m) enabled interference of multiple round trips in the cavity. The amplitude and phase of the incident field were tailored using a phase-only spatial light modulator (SLM) in combination with a spatial filter for amplitude modulation \cite{arrizon_pixelated_2007} (see Supplementary Material, Section \ref{sec:setup_ext}). The transmitted field was interfered with a reference beam in an off-axis holography configuration, and both output polarizations were imaged onto a camera. We normalized the reconstructed output field so that its total power matched the output power measured by a calibrated photodiode monitoring the transmitted light (see Supplementary Material Fig.~\ref{fig:exp_setup_extended} and Section~\ref{sec:orth_projection}). 
    
We measured the MMFC’s transmission matrix $T$ by illuminating the fiber core with a set of 242 tilted beams (121 per input polarization). The tilts were generated by applying equally spaced linear phase ramps on the SLM, spanning the angular bandwidth defined by the NA of the fiber. For each input, the fields of the two output polarizations were rearranged into a one-dimensional vector, comprising one column of the transmission matrix. To compensate for thermal drifts, we adjusted the laser wavelength by a few femtometers after every 10 input modes (see Supplementary Material, Section \ref{sec:setup_ext}, for further details on the transmission matrix measurement). The decomposition of the transmission matrix into the fiber-mode basis is depicted in Fig.~\ref{fig:2}b, exhibiting strong mode mixing with 0.8 of the energy located in the off-diagonal elements (see Supplementary Material, Section \ref{sec:fibermode_projection} for details on the fiber-mode decomposition).

\begin{figure}[ht!]
  \centering
  \includegraphics[width=0.97\columnwidth]{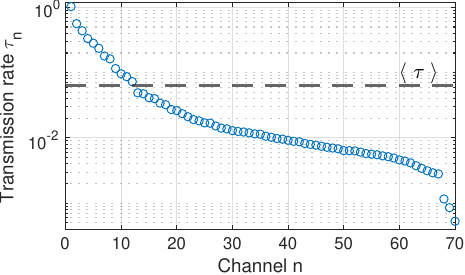}
 \caption{Transmission rates $\tau_n$ obtained by diagonalizing $T^{\dagger}T$ for an experimentally measured transmission matrix $T$.  Only the  70 most significant transmission rates are presented. The gap between channels 67 and 68 indicates that the MMFC supports 67 guided modes. The dashed horizontal line marks the average transmission $\langle \tau\rangle=0.064\pm0.001$. The uncertainties are in the order of $1\%$, smaller than the marker size (see Supplementary Material Section \ref{sec:error_estimation})}
\label{fig:3}
\end{figure}

The eigenvalues obtained by diagonalizing $T^{\dagger}T$ are presented in Fig.~\ref{fig:3}. We observed five channels with transmission rates higher than four times the average transmission, $\langle \tau \rangle = 0.064 \pm 0.001$. The presence of multiple channels with transmission rates exceeding $4\langle \tau \rangle$ demonstrates significant control over the incident wavefront. In contrast, partial control results in a transmission rate distribution that rapidly converges to the Marčenko-Pastur distribution, for which the maximal transmission rate for square matrices is $4\langle \tau \rangle$ \cite{popoff_measuring_2010,Goetschy_icc}. Additionally, we observed a consistent decrease in the eigenvalues up to a pronounced gap at the 67\textsuperscript{th} eigenvalue. This suggests that the MMFC supports 67 modes, which is largely in line with the measured transmission matrix of the fiber without dielectric coatings, exhibiting 34 guided modes and two leaky modes per polarization (see Supplementary Material Section \ref{sec:TO_measurement}).

\begin{figure}[t]
 \centering
 \includegraphics[width=0.95\columnwidth]{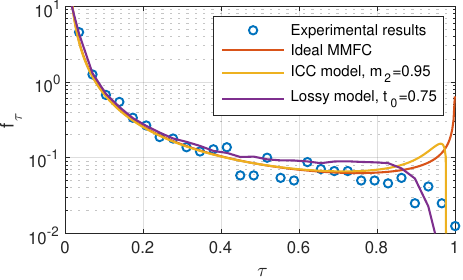}
 \caption{The probability density function (PDF) of the transmission rates, $f_\tau$, was obtained from the histogram of the transmission rates of an ensemble of 100 matrices. The experimental data (circles) is compared to the expected distribution of an ideal MMFC (red line), the incomplete channels control (ICC) model (orange line), and the lossy model (purple line). The measured data follows the ideal distribution for $\tau \lesssim 0.8$ and then gradually decays, reaching $\tau = 1$ with a finite probability. Comparison to the ICC model, which assumes full control over the input modes and control over $m_2 = 0.95$ of the output modes, demonstrates that the MMFC provides a high degree of control over both input and output modes. Additionally, comparison to the lossy model, which assumes that the four highest-order modes experience transmission of $t_0 = 0.75$ while all other modes do not experience loss, suggests that coupling to a few leaky modes can explain the deviation of the measured data from the ideal MMFC model.}
 \label{fig:hist}
\end{figure}

The transmission properties of the MMFC are governed by the statistics of the transmission rates, particularly their probability density function (PDF) $f_\tau$.
To obtain the PDF experimentally, we measured multiple realizations of the MMFC's transmission matrix by leveraging its spectral sensitivity. Specifically, we recorded 100 transmission matrices at wavelengths spaced by 20~fm, exceeding the MMFC's spectral correlation width of $\approx 10$~fm (see Supplementary Material, Section \ref{sec:wavelength_scan} and Fig.\ref{fig:wavelength_scan}). For each matrix, we computed the transmission rates and constructed their histogram (Fig.~\ref{fig:hist}). The resulting PDF remains non-negligible even for $\tau \approx 1$, reflecting the finite probability of obtaining open channels with nearly unit transmission. 

To further analyze the measured distribution of the transmission rates, we modeled the transmission through the MMFC as interference of multiple round trips within the cavity. Each round trip can be represented by $\hat{r}_1T_0^{\mathsf{T}}\hat{r}_2T_0$, where $T_0$ and $T_0^{\mathsf{T}}$ represent the transmission matrix of an uncoated fiber and its transpose, respectively, and the matrices $\hat{r}_{i=1,2}$ describe reflection from the two facets of the fiber (Fig.~\ref{fig:1}a). Analogously to a Fabry-Pérot cavity, the total transmission matrix of the MMFC is given by the infinite series of round-trip contributions (for additional details see Supplemental Material, Section \ref{sec:Bimodal}):

 \begin{figure}[b]
    \includegraphics[width=0.47\textwidth]{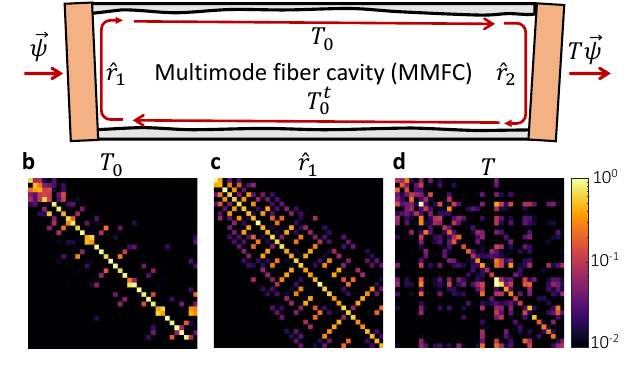}
    \caption{\label{fig:1} \textbf{Transmission matrix and mode mixing in a multimode fiber cavity (MMFC).} (a) Schematic of the MMFC: Each round trip is described by $\hat{r}_1 T_0^{\mathsf{T}}\hat{r}_2 T_0$, where $T_0$ and $T_0^{\mathsf{T}}$ are the transmission matrix of an uncoated fiber and its transpose, respectively, and $\hat{r}_1,\hat{r}_2$ denote the matrices describing the reflection from the facets. (b) Typical measured transmission matrix $T_0$ of a one-meter long uncoated stepped-index fiber supporting $N=34$ modes per polarization. The matrix exhibits weak mode coupling, with 0.8 of the energy concentrated in the diagonal blocks of the degenerate modes. (c) Numerically computed reflection matrix $\hat{r}_1$, assuming a facet angle of $\theta = 5 \times 10^{-3}$ rad,  obtained by decomposing a linear phase tilt into the fiber modes. (d) Computed transmission matrix of the MMFC using Eq.~\eqref{eq:mmfc_cavity_main}, with for $T_0$ from (b) and $r_i$ from (c).}
\end{figure}

\begin{equation}
\label{eq:mmfc_cavity_main}
\begin{split}
    T & =\hat{t}_2\frac{1}{1-T_0\hat{r}_1T_0^{\mathsf{T}}\hat{r}_2}T_0\hat{t}_1, \\
\end{split}
\end{equation}

where $\hat{t}_{i=1,2}$  represent the transmission matrices through the coated facets of the MMFC. When the facets are perfectly orthogonal to the propagation axis of the fiber, the reflection and transmission matrices remain diagonal in the fiber mode basis $\hat{r}_i=\sqrt{\rho_i} \space \mathbb{1} $, $\hat{t}_i=\sqrt{1-\rho_i} \space \mathbb{1}$. However, slight facet angles, often present in connectorized fibers \cite{senko_apex_offset}, introduce coupling between the modes, manifested by non-diagonal reflection matrices with mean reflectivity $\rho_i$ (Fig. \ref{fig:1}c).

Measurements of the transmission matrix $T_0$ for an uncoated fiber exhibit weak mode mixing, with only $0.2$ of the energy concentrated in the off-diagonal elements of the matrix (Fig.\ref{fig:1}a). In contrast, the measured transmission matrix of the MMFC shows strong mode mixing, with $0.8$ of the energy in the off-diagonal elements (Fig.\ref{fig:2}b). To investigate the origin of the higher mode mixing, we plugged into Eq.~(\ref{eq:mmfc_cavity_main}) the measured transmission matrix $T_0$, where we eliminated the effect of mode-dependent loss associated with measurement of $T_0$ by setting its singular values to unity. This computation yields a transmission matrix with $0.5$ of the energy in the off-diagonal elements. By introducing to the model a slight tilt of the fiber facets, the mode mixing is further increased. For typical tilt angles obtained in optical fibers ($\theta = 5 \times 10^{-3}$) \cite{senko_apex_offset}, mode mixing increased significantly, with $0.7$ of the energy in off-diagonal elements (Fig.~\ref{fig:1}d).

 Using the model described by Eq.~\ref{eq:mmfc_cavity_main}, incorporating mode mixing from tilted facets, we computed analytically the PDF of the transition rates within the framework of random matrix theory (see Supplementary Material Section \ref{sec:Bimodal}). As in scattering samples, it exhibits a bimodal distribution (Fig.~\ref{fig:hist}, red curve), where the exact form of the PDF depends on the coating reflectivity. For the experimental reflectivity $\rho=0.88$, the measured PDF follows the MMFC model up to transmission values of $\tau \sim 0.8$.

The slight deviation measured and model PDFs can be used to estimate the MMFC channel control, as incomplete wavefront control or losses typically reduce the number of open channels. We therefore considered two simplified single-parameter models. In the incomplete channel control (ICC) model, we assumed a lossless MMFC, while allowing for an incomplete measurement of its transmission matrix (see Supplementary Material Section \ref{sec:ICC} and Ref.\cite{Goetschy_icc}). The best agreement between the ICC model and the experimental data was obtained by assuming full control over all incident channels and 95\% control over the output channels (Fig.~\ref{fig:hist}, orange curve). In the loss model, we assumed the MMFC is lossy due to leaky fiber modes but that its transmission matrix was perfectly measured (see Supplementary Material Section~ \ref{sec:LossModel}). Consistency with the experimental data was achieved by assuming that the four highest-order modes of the fiber were leaky modes with a transmission of $t_0=0.75$, corresponding to an average loss of less than 0.015 per channel (Fig.~\ref{fig:hist}, purple curve). This assumption aligned with a direct measurement of the transmission matrix of a similar fiber without reflective coatings (see Supplementary Material, Section~\ref{sec:TO_measurement}). Neither model perfectly fits the data, but both suggest high channel control and low overall loss.

In scattering media, open transmission eigenchannels exhibit significantly longer dwell times compared to the average value \cite{rotter2017light,durand2019optimizing}. To investigate whether this property of open channels also holds in MMFCs, we performed numerical simulations of light propagation in our MMFC setup and computed the expectation values of the dwell time operator, $Q=- i \left( T^\dagger \frac{dT}{d\omega} + R^\dagger \frac{dR}{d\omega} \right)$, where $R$ is the reflection matrix of the MMFC (see Supplementary Material Section~\ref{sec:dwell_time} for details) \cite{durand2019optimizing}. The computation reveals that the dwell time of the open channel (110 ns) exceeds the mean dwell time (7 ns) by more than an order of magnitude and that the dwell time of transmission eigenchannels increases monotonically with their transmission rates (see Fig. \ref{fig:lossy_MMFC_model}).

\begin{figure}
    \centering
    \includegraphics[width=1\columnwidth]{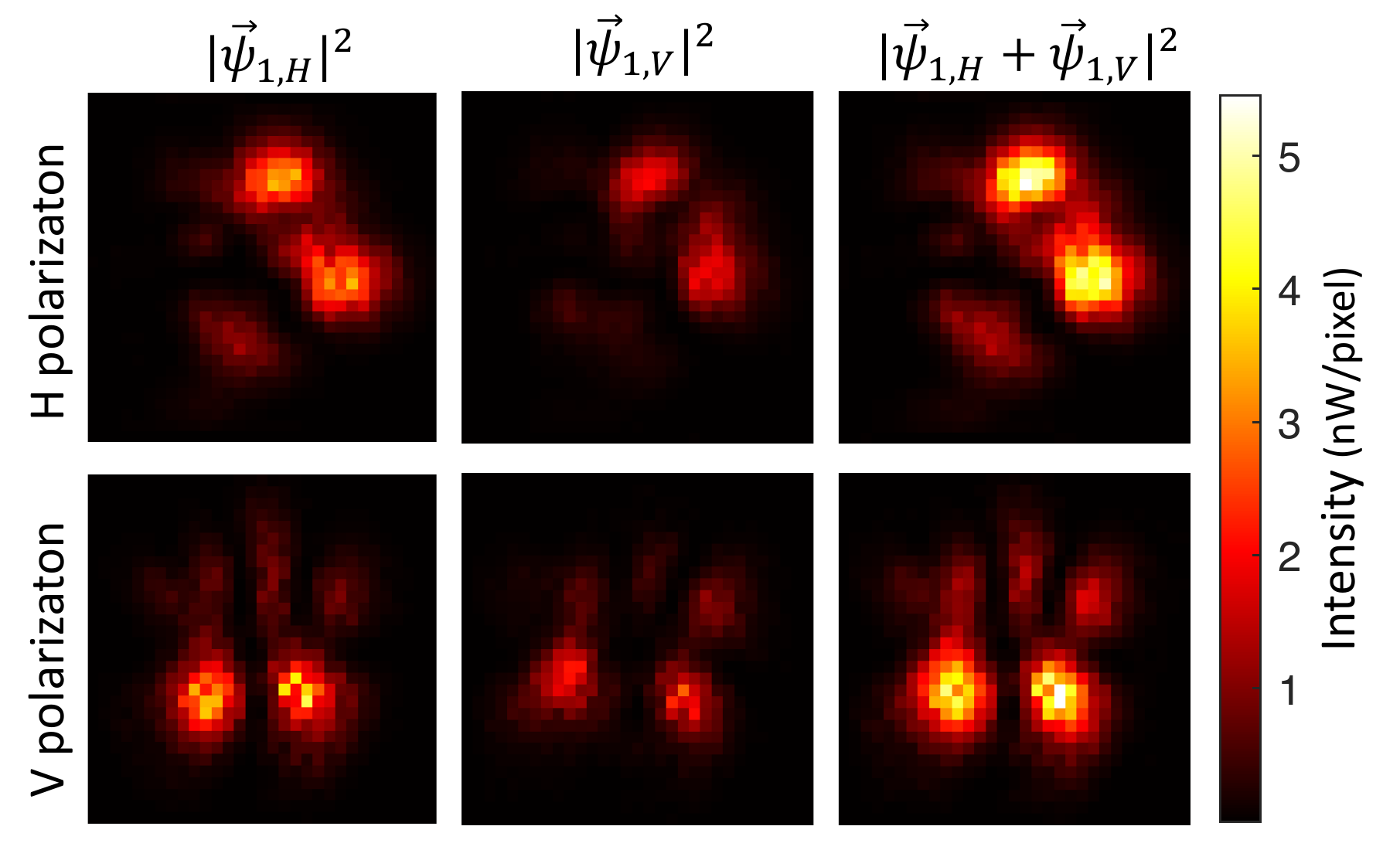}
    \caption{Selective excitation of the highest transmission channel. The first transmission eigenchannel $\vec{v}_1$ was decomposed into its two input polarization components, $\vec{v}_{1,H}$ and $\vec{v}_{1,V}$, which were excited sequentially. The left column shows the measured output intensity, $|\vec{\psi}_{1,H}(x,y)|^2$, for excitation of the horizontal component, while the central column shows the output intensity, $|\vec{\psi}_{1,V}(x,y)|^2$, for excitation of the vertical component. The right column depicts the intensity for the coherent superposition, $|\vec{\psi}_{1,H}(x,y) + \vec{\psi}_{1,V}(x,y)|^2$. The top row represents the horizontal polarization component of the output, and the bottom row represents the vertical polarization component. The output intensity patterns were normalized by the power measured at the input of the MMFC, ensuring that the sum over all pixels corresponds to the transmission rate of the pattern. The transmission rates for $\vec{\psi}_{1,H}$, $\vec{\psi}_{1,V}$, and $\vec{\psi} = \vec{\psi}_{1,H} + \vec{\psi}_{1,V}$ are $\tilde{\tau}_{1,H} = 0.60$, $\tilde{\tau}_{1,V} = 0.35$, and $\tilde{\tau}_{1} = 0.90$, respectively.  
    }
    \label{fig:excitation}
\end{figure}

The MMFC platform’s high control enables selective excitation of transmission eigenchannels and study of their spatial structure. To excite an open channel of the system, we tailored the incident wavefront to match the first transmission eigenchannel of the transmission matrix, namely the input channel $\vec{v}_1$ with the highest transmission rate $\tau_1$. Since mode mixing in the MMFC is polarization-dependent (see Fig.~\ref{fig:2}b), the incident field corresponding to $\vec{v}_1$ consisted of two different polarization components. However, as the SLM  can shape only one polarization at a time, we first sent the horizontal polarization component $\vec{v}_1$ into the fiber and measured the output field  $\vec{\psi}_{1,H}$. At the output, both polarization components were measured simultaneously, so $\vec{\psi}_{1,H}$ includes both components, with the subscript $H$ indicating the input polarization. Next, we rotated the input polarization state by $90^\circ$, sent the vertical component of $\vec{v}_1$, and measured the output field $\vec{\psi}_{1,V}$. Finally, we coherently combined the measured output fields to obtain the total output field for simultaneous excitation of both input polarizations: $\vec{\psi}_{1} = \vec{\psi}_{1,H} + \vec{\psi}_{1,V}$, as depicted in Fig.~\ref{fig:excitation}.

To directly measure the transmission rate of the excited open channel, we normalized the total output power, $\|\vec{\psi}_{1,H}+\vec{\psi}_{1,V}\|^2$, by the total input power, $p_H + p_V$, measured with a calibrated photodetector monitoring the input beam before entering the MMFC (Supplementary Fig.~\ref{fig:exp_setup_extended}) . The measured transmission rate is given by $\tilde{\tau}_1 = \frac{\| \vec{\psi}_{1,H}+\vec{\psi}_{1,V}\|^2}{p_H + p_V}.$ We obtained $\tilde{\tau}_1 = 0.90 \pm 0.04$, representing an 18-fold enhancement compared to the average transmission measured for a set of random inputs. The measured transmission $\tilde{\tau_1}$ is about $10\%$ lower than the transmission rate $\tau_1$ computed from the transmission matrix, but the 18-fold enhancement relative to the average intensity remains consistent in both cases. This discrepancy likely arises from inaccuracies in the transmission matrix measurement and phase instabilities.

Figure~\ref{fig:excitation} shows that the spatial distribution of the output fields $\vec{\psi}_{1,H}$ and $\vec{\psi}_{1,V}$ are highly correlated, despite their orthogonal input polarizations. This spatial overlap enables efficient constructive interference, which is required for open channels with near-unity transmission rates. Similarly, multiple round trips in the cavity are expected to interfere constructively at the fiber output. In a perfect fiber without mode mixing, guided modes maintain their transverse shape and open channels correspond to the guided modes whose propagation constants satisfy constructive interference at the output of the fiber. However, in the presence of mode mixing, none of the modes retain their transverse shape preventing good spatial overlap of multiple round trips. Indeed, when we coupled the guided modes of the fiber (LP$_{lm}$) to the MMFC, the highest transmission rate we observed was $\tilde{\tau}_{LP}=0.17$, far below $\tilde{\tau}_1$. MMFC open channels uniquely retain transverse shape across round trips despite mode mixing (see Supplementary Material Section \ref{sec:Bimodal}). While we cannot directly probe the transverse shape after each round trip, we observed indirect evidence of this feature by tuning the wavelength of the laser by a few femtometers. For open channels, the intensity quickly dropped, while the transverse shape stayed constant. This suggests that the wavelength detuning was too small to change the interference between the modes, resulting in similar output intensity patterns, but large enough to change the relative phase between different round trips.

In this work, we investigated open channels and their distribution in MMFCs. By selectively exciting an open channel,  we achieved a transmission rate of $0.90 \pm 0.04$, corresponding to an 18-fold enhancement over the measured average transmission rate $\langle \tilde{\tau} \rangle = 0.049 \pm 0.001$. To gather statistical insights, we measured multiple transmission matrices at different wavelengths and extracted the PDF of the transmission rates. The obtained PDF exhibited a nonzero probability of channels with $\tau = 1$, though slightly lower than predicted by the ideal bimodal distribution. This deviation allowed us to estimate the degree of control and loss in our system, suggesting that we controlled over 0.95 of the MMFC modes.

MMFCs enable high control by confining channels with a finite angular spread, making them easily addressable with an SLM. This control revealed strong correlations between orthogonal input polarizations in near-unit transmission channels, highlighting their origin in optimal constructive interference and long dwell times.

MMFCs offer exceptional scalability, with the number of channels readily extendable to thousands by increasing the core size and numerical aperture. This scalability, combined with long dwell times, compatibility with extended fiber lengths, and strong confinement that enables operation in the nonlinear regime, positions MMFCs as a powerful platform for studying wave transport and nonlinear dynamics in complex media.

\begin{acknowledgments}
We wish to thank Hui Cao and Valentin Freilikher for insightful and fruitful discussions. Y.B. acknowledges support from Israeli Science Foundation Grants No. 1363/15 and 2497/21, and the Zuckerman STEM Leadership Program. A.G. and S.M.P. acknowledge support by the program ``Investissements d'Avenir'' launched by the French Government.

\end{acknowledgments}

\bibliographystyle{apsrev4-2}
\bibliography{refs}

\input{SI.tex}

\end{document}

%% file: SI.tex
\def\thefigure{S\arabic{figure}}
\setcounter{figure}{0}
\renewcommand{\theequation}{S\arabic{equation}}
\setcounter{equation}{0}

\section{Supplementary Information}

\subsection{Additional experimental details}
\label{sec:setup_ext}
\begin{figure}[b!]

 \centering
 \includegraphics[width=1\columnwidth]{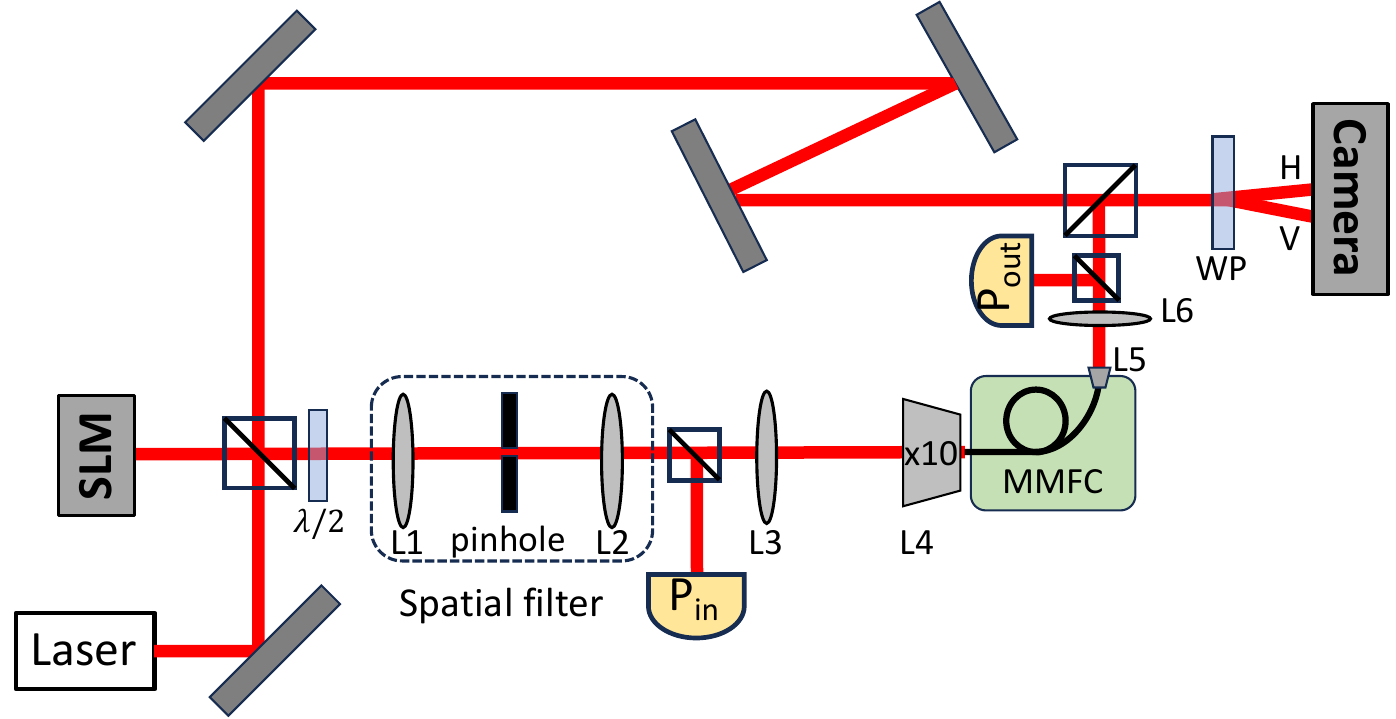}
 \caption{Extended setup. The multimode fiber cavity (MMFC) is placed in the sample arm of a Mach Zehnder interferometer. The sample arm also includes a reflecting SLM that is reimaged on the input facet of the MMFC with 4 lenses (L1-L4) in an 8f configuration with a demagnification of x100. A pinhole placed between lenses L1 and L2 is used for spatial filtering, converting the phase-only SLM into an amplitude and phase modulator. A half waveplate ($\lambda/2$) mounted on a motorized rotational stage is used to rotate the polarization of the input field. The output facet of the MMFC is imaged onto a CMOS camera using two additional lenses (L5,L6), and a Wollaston prism (WP) is used to measure the two output polarizations simultaneously.  The input and output powers are measured with two calibrated photodiodes ($P_\mathrm{in}$,$P_\mathrm{out}$).} 
 \label{fig:exp_setup_extended}
\end{figure}

Figure \ref{fig:exp_setup_extended} depicts the experimental setup. We used a single frequency extended cavity diode laser (Toptica DLC-Pro 632) with a long coherence length ($>$ 100 m) to allow interference of multiple round-trips in the 1m long MMFC. Fine-tuning of the laser's wavelength at the 10-femtometer scale was used to compensate for thermal drifts of the MMFC. For the wavefront shaping, we used an LCOS spatial light modulator in reflecting mode (Holoeye PLUTO-NIR-011). The SLM is reimaged onto the input facet of the MMFC using four lenses with focal lengths of $f_1=400$mm, $f_2=45$mm, $f_3=200$mm, and $f_4=18$mm, with a total demagnification of $\approx100$.  Using a pinhole at the Fourier plane between lenses L1 and L2, we blocked light outside of the first diffraction order of the phase pattern applied by the SLM, allowing phase and amplitude modulation from a phase-only SLM \cite{arrizon_pixelated_2007}. Since the SLM can modulate only a single polarization, we used a $\lambda/2$ waveplate on a motorized rotating stage to tailor the wavefront of each polarization component sequentially. For each polarization input, the complex output field is measured and the output field for simultaneous excitation of both input fields is then computed. 

The output of the MMFC is reimaged onto a CMOS camera (Mako U-130B) with two lenses with focal lengths $f_5=18.24$mm and $f=750$mm, and interfered with the light coming from the reference arm. A Wollaston prism placed before the camera is used to measure two orthogonal polarization states simultaneously. The prism is rotated by $45\deg$ to measure the $\pm45\deg$ polarization components, avoiding polarization rotation of the reference arm. 

To directly measure the transmission rates, we use two photodiodes that monitor the power before and after the MMFC ($p_\mathrm{in}$ and $p_\mathrm{out}$ in Figure \ref{fig:exp_setup_extended}). The photodiodes are calibrated with the same powermeter that was placed right after L4 for $p_\mathrm{in}$ and after L5 for $p_\mathrm{out}$. Note that since we measure the output power after the fiber collimator L5, we measure a slightly lower power than the power right at the output of the MMFC due to reflections, yet since the collimator is coated with anti-reflection coating we underestimate the output power by a fraction of a percent.

To reduce the thermal drift of the fiber, we place the MMFC in a plastic box and glued the fiber to an aluminium plate. Residual thermal drifts are observed during the 5 minutes it takes to measure a transmission matrix of the MMFC. To quantify the drift, we measured for a fixed input field the correlation between the output fields measured at different times. The correlation drops to 0.5 after $\approx1$ minutes, while the measurement of a single TM takes 5 minutes. To compensate for the thermal drift, the wavelength of the laser can be slightly tuned, as to first-order thermal expansions of fibers can be compensated by wavelength tuning. To this end, during the measurement of a transmission matrix, every few seconds we send the same input field, and tune the wavelength of the laser to maximize the correlation between the output field and the field recorded at the beginning of the measurement. With slight wavelength adjustments of a few femtometers, we could keep the correlation above 0.98 for over 30 minutes, much longer than the time required for our measurements.

\subsection{Projection of the measured transmission matrix on an orthonormal input basis}
\label{sec:orth_projection}
The physical meaning of the transmission rates $\tau_n$, defined by eigenvalues of the matrix $T^\dagger T$, is often associated with the fraction of the power that is transmitted through the sample when the input field matches the corresponding eigenvector $\vec{v}_n$. This is true, however, only when the input modes that were used to measure the transmission matrix $T$ are orthogonal \cite{pai2020optical}.  In practice, the input modes are often non-orthogonal since it is desirable to over-sample the transmission matrix and measure it with more input modes than the number of guided modes the fiber supports.  One therefore needs to project the measured transmission matrix onto an orthonormal basis before computing its transmission rates. 

To find the desired projection, we first write the incident fields in a matrix form. We denote by the matrix element $S_{i,j}$  the field of the $i^{th}$ incident mode at the transverse coordinate $(x_j,y_j)$ on the input facet of the fiber 
$S_{i,j}=\text{circ}\left(\frac{\sqrt{x_j^2+y_j^2}}{R_{core}}\right)e^{\frac{2\pi}{\lambda}(\theta^{(i)}_xx_j+\theta^{(i)}_yy_j)}$. The function $\text{circ}(x)=1$ for $0\leq x\leq 1$ and 0 otherwise, represents the binary mask applied with the SLM to block light outside the core radius $R_{core}$. Due to the finite extent of the incident fields, the columns of the matrix $S$ are not orthogonal and $S$ is not unitary. 

The measured transmission matrix $\tilde{T}$ is related to the true transmission matrix of the MMFC $T$ by $\tilde{T}=TS$. The columns of the matrices $\tilde{T},S$ are normalized so that the sum of the absolute value squared of each column of $\tilde{T}$ ($S$) equals the power measured by the photodiode $p_\mathrm{out}$~$(p_\mathrm{in})$.

Since $S$ is not unitary, the transmission rates of $\tilde T$ are different than the transmission rates of $T$. We, therefore, need to compute the transmission rates of $T=\tilde{T}S^{-1}$. To invert the matrix $S$, we first compute its singular value decomposition $S=U_s\Sigma V_s^\dagger$ and then compute the pseudoinverse of the $\Sigma$ by inverting all the diagonal elements above some threshold. Finally, we compute the transmission rates of:

\begin{equation}
T=\tilde{T}S^{-1}=\tilde{T}V_s\Sigma^{-1}U^{\dagger}_s
\end{equation}

In practice, the multiplication by the matrix $U^{\dagger}_s$ does not change the transmission rates of $T$ since it is a unitary matrix. For the same reason, in this analysis, we do not consider aberrations that are also modeled by a unitary matrix that does not change the transmission rates.  

\subsection{Projection of the measured transmission matrix on the fiber mode basis}
\label{sec:fibermode_projection}

Optical fiber characterization is highly sensitive to optical aberrations and misalignments. 
Although such effects do not alter the transmission rate or their distribution, 
they complicate the projection of the TM onto the fiber mode basis, which is essential for a physical interpretation of the measurements. In particular, 
the estimation of mode-dependent losses and mode coupling is highly susceptible to even minor modifications of the system~\cite{Ploschner2015}. 
To mitigate measurement degradation, 
we numerically identify and correct 
for these aberration effects using the approach in \cite{Maxime:19}. 
This procedure entails learning the input and output aberrations by optimizing a model-based numerical model 
leveraging deep learning frameworks. 
The solver aims to minimize the energy loss 
when projecting the measurements onto the known mode basis 
of the fiber. 
Since all the energy transmitted through the optical system 
must be conveyed by the fiber modes, 
accurately accounting for aberrations and misalignments 
ensures that projection onto the mode basis does not alter the energy of the TM. 
This method facilitates the acquisition of an accurate TM of the system, even with an imperfectly calibrated optical setup.

\subsection{Uncertainty estimations for transmission eigenvalues}
\label{sec:error_estimation}
To estimate the uncertainty of the measured transmission rates, we added artificial noise to the measured transmission matrices and computed the standard deviation of the observed transmission rates for a large number of noise realizations. We chose a simple noise model that we believe captures the main noise source in our measurements, excess intensity noise of the measured powers, which we estimate by 3\%. 

To numerically estimate the uncertainty in the transmission rates, we performed the following steps: 
(a) Measured a transmission matrix.
(b) Multiplied and divided each output vector by the square root of a random number distributed normally with a mean of 1 and a standard deviation of 0.03, to take into account both input and output power uncertainties.
(c) Computed the transmission rates of the matrix with the artificial noise
(d) Repeated steps (b-c) 100 times and computed the mean and standard deviation of each transmission rate. We find that the standard deviation of the high transmission rates is 0.01.

\subsection{Experimental characterization of the uncoated fiber}
\label{sec:TO_measurement}
\begin{figure*}[hbt!]
 \centering
 \includegraphics[scale=0.51]{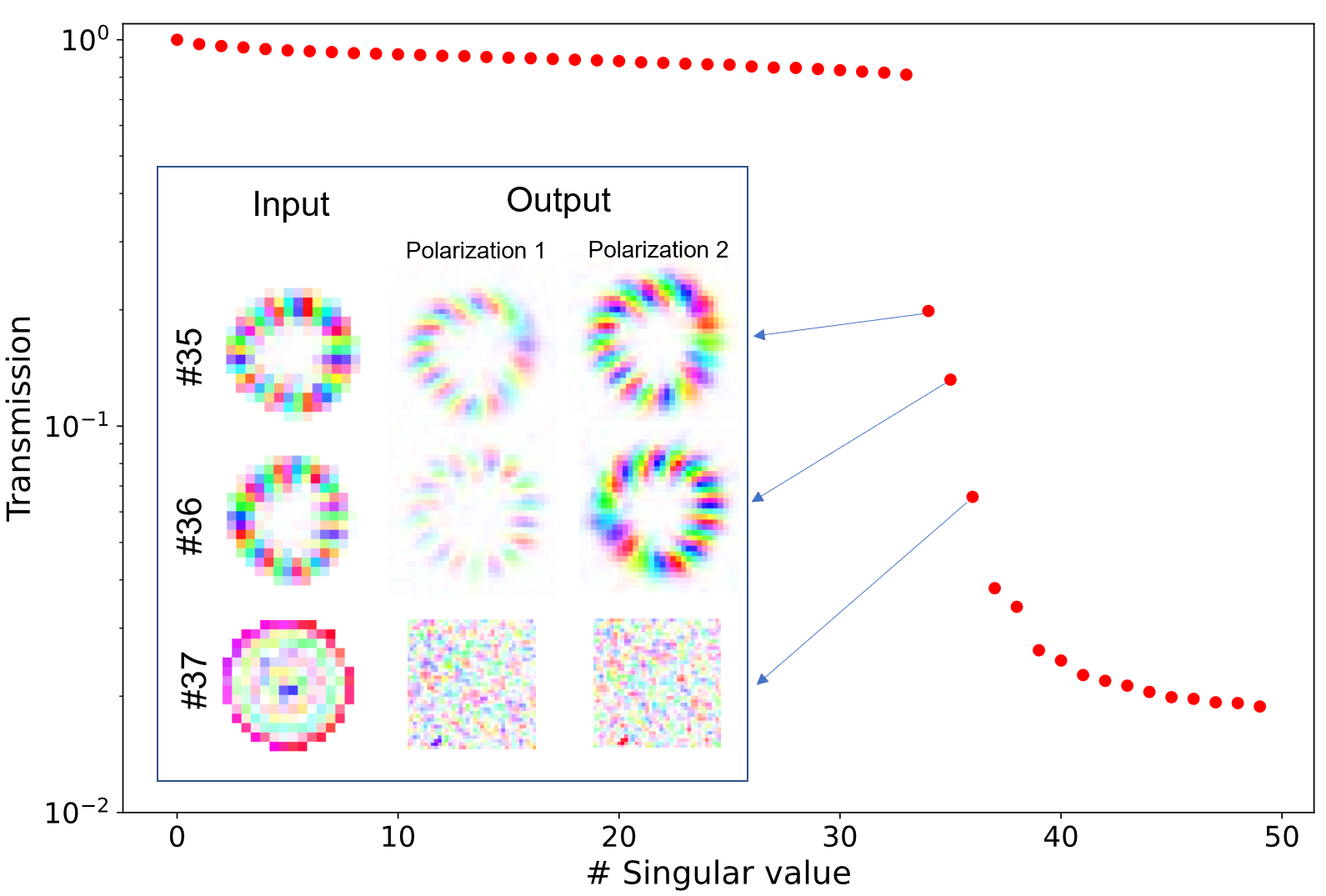}
 \caption{Transmission rates and leaky modes obtained from a measured single-polarization transmission rate of an uncoated multimode fiber. The first 34 channels exhibit high transmission rates ($>$0.8) followed by a sharp drop. The inset depicts the complex field of the input and output patterns of channels 35, 36 and 37.  The output fields of channels 35 and 36 match the patterns of a guided mode of the fiber. The output field of channel 37 is clearly noise, and its input field is concentrated in the clad of the fiber.  We therefore conclude that for each polarization, the uncoated fiber supports 34 guided modes and two leaky modes.}
  \label{fig:lossy_MMF}
\end{figure*}

To check if our MMFC contains lossy channels, we took a non-coated multimode fiber with identical parameters to the fiber we used for the MMFC, and measured its transmission matrix for a single input polarization.
The results depicted in Fig. \ref{fig:lossy_MMF} clearly show 34 guided modes, followed by two leaky modes. 
The transmission rate of the 37\textsuperscript{th} channel is on the order of the transmission rates of the leaky modes (35 and 36), yet inspection of its input intensity pattern reveals that its energy is mostly concentrated in the clad and not in the core (inset of Fig.~\ref{fig:lossy_MMF}). Moreover, since the spatial distribution of it output field is indistinguishable from noise, we conclude it is not a guided mode of the fiber. We therefore model the two-polarization transmission matrix of the MMFC with 72 guided modes, including four lossy channels.

\subsection{Spectral dependence of the output field}
\label{sec:wavelength_scan}
\begin{figure}[t]
 \centering
 \includegraphics[width=1\columnwidth]{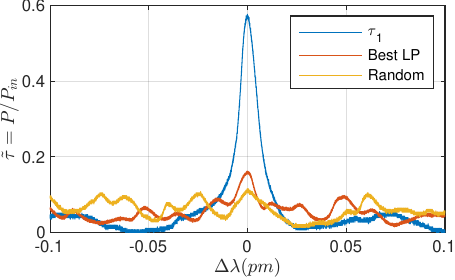}
 \caption{ Spectral dependence of output fields. The measured transmission rate $\tilde\tau$ is defined by the ratio of the output and input powers as we scan the wavelength for different input fields: a highly transmitting channel, the best transmitted LP mode and a random input. The highly transmitting channel has a peak transmittance of 0.57 at the wavelength where the TM was measured. It then quickly drops to the mean transmission of 0.05 for a wavelength detuning of 10fm, indicating that the eigenchannels are wavelength dependent.
 }
 \label{fig:wavelength_scan}
\end{figure}

By tuning the wavelength of the laser, we studied the spectral dependence of the output field for various inputs. We compared the outputs for three different input fields: (i) the horizontal input polarization component of a channel with a high transmission rate, obtained from a transmission matrix measured at wavelength $\Delta\lambda=0$; (ii) the fiber’s LP mode that yields the highest transmission rate; and (iii) a random superposition of input modes. We scanned a bandwidth of $\pm 0.2$ pm around $\Delta\lambda=0$ and recorded the transmitted power for each wavelength (see Fig. \ref{fig:wavelength_scan}). 

The open channel with the highest transmission rate showed a sharp peak at $\Delta\lambda=0$, with a transmission rate of 0.57 (an 11-fold enhancement). As we slightly detuned the wavelength, the transmission decreased rapidly.  With a small wavelength detuning of $\Delta\lambda=\pm10$ fm, the transmission rate dropped to 0.17 (a 3-fold enhancement), matching the peak transmission of the best LP mode. For the random input, we observed fluctuations around an average transmission of 0.05.

\subsection{Theoretical analysis of the MMFC}
\label{sec:distr}
In this section, we derive the transmission eigenvalue distribution of MMFCs using random matrix theory (RMT).  We then show how this distribution is modified when the input and output spaces of the transmission matrix are partially measured, using the \textit{incomplete channel control} (ICC) model. Finally, we discuss an alternative model that includes loss during propagation.

\subsubsection{The bimodal probability density function (PDF)}
\label{sec:Bimodal}

We consider a reflectionless, non-absorbing multimode fiber described by a random unitary matrix $T_0$, which supports $N$ propagating channels and is placed between two partial reflectors characterized by the reflection and transmission matrices $\hat{r}_1$, $\hat{t}_1$ and $\hat{r}_2$, $\hat{t}_2$, respectively. By summing all cavity round trips contributing to the transmission of the system, we obtain the following expression for the total transmission matrix:
\begin{equation}
\label{eq:mmfc_cavity}
\begin{split}
    T  = &\hat{t}_2T_0\hat{t}_1 + \hat{t}_2(T_0\hat{r}_1T_0^\mathsf{T}\hat{r}_2)T_0\hat{t}_1 \\
       &+ \hat{t}_2(T_0\hat{r}_1T_0^\mathsf{T}\hat{r}_2)^2T_0\hat{t}_1 + \dots \\
       =&\hat{t}_2\frac{1}{1- T_0 \hat{r}_1 T_0^\mathsf{T}\hat{r}_2} T_0\hat{t}_1.
\end{split}
\end{equation}

Here, $T_0$ describes the transmission from the proximal end to the distal end of an uncoated fiber, while the transmission from the distal end to the proximal end is described by $T_0^\mathsf{T}$ (the transpose of $T_0$), as dictated by optical reciprocity. 

To access the eigenvalues of the matrix $T^\dagger T$, we can take advantage of the singular value decompositions of the reflection and transmission of the partial reflectors. For simplicity, we assume that all reflection singular values of the reflectors are identical and equal to $\sqrt{\rho}$. However, we include the possibility of mode mixing. Using the constraints imposed by current conservation and time-reversal symmetry, we write 
\begin{equation}
\begin{split}
\hat{t}_1&=U_1\sqrt{1-\rho}V_1^\dagger, \;\; \hat{r}_1=U_1\sqrt{\rho}U_1^\mathsf{T}, 
\\
\hat{t}_2&=U_2\sqrt{1-\rho}V_2^\dagger, \;\; \hat{r}_2=-V_2^*\sqrt{\rho}V_2^\dagger,
\end{split}
\end{equation}
where $U_1, V_1, U_2, V_2$ are unitary matrices that are non-diagonal in the case of mode mixing. 

With this decomposition, the matrix $T^\dagger T$ can be expressed as
\begin{equation}
    \label{eq:TdagT}
    T^\dagger T=V_1U_1^\dagger T_0^\dagger \left[ a +\frac{b}{2}(\Omega+\Omega^\dagger) \right] ^{-1}T_0 U_1V_1^\dagger,
\end{equation}
where $\Omega$ is a matrix describing a round trip in an uncoated fiber:
\[
\Omega=T_0 U_1 U_1^\mathsf{T}T_0^\mathsf{T}V_2^*V_2^\dagger,
\]
and the parameters $a$ and $b$ quantify the reflectivity of the coatings:
\begin{equation}
\begin{split}
    a & =\frac{1+\rho^2}{(1-\rho)^2},\\
    b & =\frac{2\rho}{(1-\rho)^2}.
\end{split}    
\end{equation}

Since $T_0$ is unitary, the eigenvalue spectrum of $T^\dagger T$ matches that of $\left[ a +\frac{b}{2}(\Omega+\Omega^\dagger) \right] ^{-1}$. Hence, we can express the transmission eigenvalues $\tau_n$ of $T^\dagger T$ in terms of the eigenvalues $\lambda_n=e^{i\phi_n}$ of the unitary matrix $\Omega$:
\begin{equation}
\label{eq:tau_n}
    \tau_n=\left[a+b\cos(\phi_n)\right]^{-1}.
\end{equation}

Moreover, for a fiber length much longer than the wavelength, the phases $\phi_n$, which correspond to the phase accumulated along a round trip between the two reflectors, are assumed to be uniformly distributed in $\left[0,2\pi\right]$. Therefore, the transmission eigenvalue PDF takes the form
\begin{equation}
\label{eq:distr}
\begin{split}
    f_\tau&=\int \frac{d\phi}{2\pi}\delta(\tau-[a+b\cos(\phi)]^{-1}) \\
           &= \frac{1}{\pi}\frac{\sqrt{\tau^-}}{\tau}\frac{1}{\sqrt{(\tau-\tau^-)(1-\tau)}},
\end{split}
\end{equation}
where
\begin{equation}
    \tau^-=\left(\frac{1-\rho}{1+\rho}\right)^2.
\end{equation}

The distribution in Eq.~(\ref{eq:distr}) has a bimodal shape in the interval $\left[\tau^-,1\right]$, with a mean value $\langle\tau\rangle=\sqrt{\tau^-}$ (Fig.~\ref{fig:ICC_theory}, green curve). A similar bimodal distribution was derived to describe the resistance of a double barrier junction in Ref.~\cite{melsen1995conductance}.  

The probability $P_1$ of having at least one channel with transmission higher than $\tau_c$ in a system with $N$ channels is therefore $P_1=1-p^N$, where $p=\int_{\tau^-}^{\tau_c} f(\tau) d\tau$ is the complementary probability of all $N$ channels having transmission rates below $\tau_c$. For $\tau_c=0.9$ and $N=34$, we find $P_1=0.6$.

To gain intuition into the mechanism that yields open channels with transmissions close to unity, let us consider a channel with $\tau_n=1$. Inspection of Eq.~(\ref{eq:tau_n}) reveals that this channel corresponds to an eigenmode of the round-trip matrix $\Omega$ with an eigenvalue $\lambda_n=-1$. This fact is also apparent directly from Eq.~(\ref{eq:mmfc_cavity}). Therefore, we conclude that open channels of a multimode fiber cavity with $\tau=1$ correspond to modes that preserve their transverse shape after each round trip. Different round trips in the cavity thus overlap spatially, yielding efficient constructive interference of the transmitted fields and perfect destructive interference of the fields reflected from the input facet.

\subsubsection{Transmission eigenvalue PDF with incomplete channel control (ICC)}
\label{sec:ICC}
\begin{figure}[h!]
\begin{centering}
\includegraphics[width=1\columnwidth]{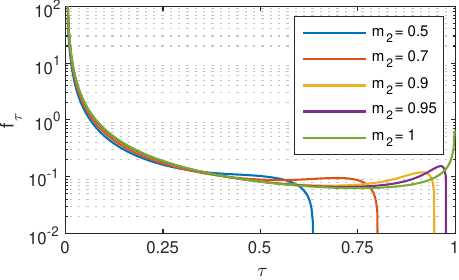}
    \caption{\textbf{The probability density function of the transmission rates of an MMFC} with reflectivity $\rho=0.88$ (green curve). Incomplete control significantly lowers the probability of obtaining open channels with near-unity transmission. Even for full input control ($m_1=1$) and nearly full control over the output modes ($m_2\simeq0.95$, purple curve), the probability of getting $\tau=1$ vanishes.
    }
    \label{fig:ICC_theory}
    \end{centering}
\end{figure}

In practice, it is very difficult to achieve full control over all the input and output modes of the system. Hence, it is essential to consider the expected distribution of the transmission rates in the case of incomplete channel control (ICC). Let us assume the accessible transmission matrix $\tilde{T}$ of the system is of size $M_2\times M_1$, with $M_1,M_2\leq N$. We can model it as a filtered random matrix drawn from the larger $N\times N$ matrix $T$. The mathematical relation between the PDFs of $\tilde{T}^\dagger \tilde{T}$ and $T^\dagger T$ has been established in Ref. \cite{Goetschy_icc}. It is given by
\begin{equation}
    \label{eq:icc_sol}
    \tilde{f}_\tau=\lim_{\eta\rightarrow0^+} \frac{1}{\pi} \Im[\tilde{g}(\tau+\mathrm{i}\eta)],
\end{equation}
where $\tilde{g}(z)$ is the solution of the implicit equation
\begin{equation}
\label{eq:icc_implicit}
    N(z)g[N(z)^2/D(z)]=D(z).
\end{equation}
Here $N(z)$ and $D(z)$ are two auxiliary functions defined as
\begin{equation}
    N(z)=zm_1\tilde{g}(z)+1-m_1,
\end{equation}
\begin{equation}
    D(z)=m_1\tilde{g}(z)\left[zm_1\tilde{g}(z)+m_2-m_1\right],
\end{equation}
with $m_1=M_1/N$ and $m_2=M_2/N$. The function $g(z)$ is the Stieltjes transform of the PDF given in Eq.~(\ref{eq:distr}),
\begin{equation}
    \begin{split}
        g(z)&=\int_{\tau^-}^1 d\tau \frac{f_\tau}{z-\tau} \\
            &= \frac{1}{z}\left[1+i\frac{\sqrt{\tau^-}}{\sqrt{(z-\tau^-)(1-z)}}\right].
    \end{split}
\end{equation}
We solve for $\tilde{g}(z)$ and present the results of Eq.~(\ref{eq:icc_sol}) in Fig. \ref{fig:ICC_theory}, where for full control $m_1=m_2=1$, the solution is $\tilde{g}(z)=g(z)$, and we recover the original bimodal distribution (Eq.~( \ref{eq:distr})). It is worthwhile to mention that even a small degree of ICC will suppress the highly transmitting eigenchannels. 

\subsubsection{The lossy MMFC model}
\label{sec:LossModel}

Next, we numerically computed the transmission rate distribution for a lossy MMFC. This was done by sampling a random orthogonal matrix $T_0$ and setting four of its eigenvalues to be lowered from 1 to $\sqrt{t_0}\in[0,1]$. We computed the transmission matrix of the MMFC using Eq.~(\ref{eq:mmfc_cavity}). 
We repeated this computation 5000 times for different realizations of $T_0$. The results for various loss values for the leaky modes $t_0$ are shown in Fig. \ref{fig:lossy_MMFC_model}. 
As expected, without loss ($t_0=1$), we obtained the original bimodal distribution (blue curve). 
Interestingly, the peak at $\tau=1$ disappears as early as $\tau_0=0.9$, when only four modes have a transmission of 0.9, while the rest have a transmission of 1. 

\begin{figure}[b!]

 \centering
 \includegraphics[width=0.95\columnwidth]{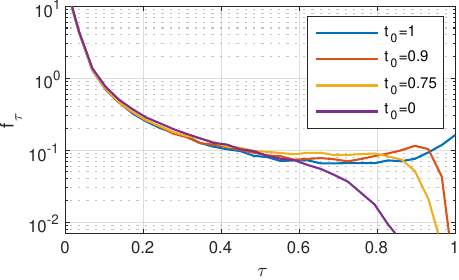}
 \caption{Lossy MMFC model. Numerical results for the transmission rate distribution of an MMFC with loss. The loss model assumes the transmission of four of the 72 fiber modes is $t_0<1$. We computed the transmission rates for 5000 numerically generated random matrices following this loss model. The open channel peak at $\tau=1$ disappears even for small losses in the fiber ($t_0=0.9$).}
 \label{fig:lossy_MMFC_model}
\end{figure}

\subsection{Numerical computation of the dwell time in the MMFC}
\label{sec:dwell_time}

In scattering media, open channels are associated with increased dwell times within the sample. To investigate whether this holds in MMFCs, we computed the dwell-time operator \cite{durand2019optimizing}:  
\begin{equation}
\label{eq:DwellTime}
Q=- i \left( T^\dagger \frac{dT}{d\omega} + R^\dagger \frac{dR}{d\omega} \right),
\end{equation}
where the transmission and reflection matrices of the MMFC, $T$ and $R$, are respectively defined by Eq.~\eqref{eq:mmfc_cavity} and 
\begin{equation}
\label{eq:reflection}
R = \hat{r}'_1 + \hat{t}_1^\mathsf{T} T_0^\mathsf{T} \hat{r}_2 T_0 \frac{1}{(1 - \hat{r}_1 T_0^\mathsf{T} \hat{r}_2 T_0)} \hat{t}_1,
\end{equation}
where $\hat{r}'_1 = -V_1^*\sqrt{\rho}V_1^\dagger$ represents the reflection matrix of the first facet for waves incident from the left.

For $T_0$, we assume a perfect fiber with no mode mixing, described by the diagonal matrix $T_0^{(m,n)}= e^{i\beta_m(\omega)L} \delta_{mn}$, where $\beta_m(\omega)$ is the frequency-dependent propagation constant of the $m$th mode in a perfect fiber. Since for a fiber of length $L = 1$ m, the spectral scale of variation of $T_0$ is $\frac{\Delta\omega}{\omega} \approx 10^{-6}$, we can neglect the spectral dependence of both the number of modes and their spatial profiles. 
Numerical computation of Eq.\eqref{eq:DwellTime} for a model that corresponds to the MMFC parameters in our experiment reveals that the dwell time of an open channel with transmission rate $\tau=0.92$ is $110$ns, an order of magnitude longer than the average dwell time of $7$ns. Figure \ref{fig:lossy_MMFC_model} shows the dwell time as a function of the transmission rate for all eigenchannels of the  transmission matrix, demonstrating that higher transmission rates are associated with longer dwell times.

\begin{figure}[b!]

 \centering
 \includegraphics[width=0.95\columnwidth]{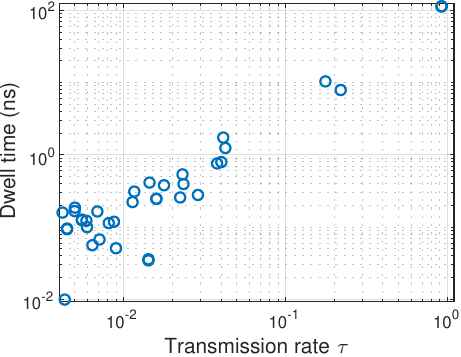}
 \caption{Dwell times of the singular transmission channels of the MMFC transmission matrix. Numerical results for the dwell times in the MMFFC, computed using the expectation values of the dwell-time operator Q. The MMFC parameters assumed for this computation were reflectivity $\rho=0.88$, fiber length $L=1$ m, and facet angle $\theta=5\times 10^{-3}$.}
 \label{fig:lossy_MMFC_model}
\end{figure}